\documentclass{aa}  

\usepackage{graphicx}
\usepackage{txfonts}
\begin{document}

   \title{Uncovering new white dwarf--open cluster associations using \textit{Gaia} DR3}


   \author{M. Pri{\v s}egen \inst{1,2} \and N. Faltov\'{a} \inst{2}}

   \institute{Advanced Technologies Research Institute, Faculty of Materials Science and Technology in Trnava, Slovak University of Technology in Bratislava, Bottova 25, 917 24 Trnava, Slovakia \and Department of Theoretical Physics and Astrophysics, Faculty of Science, Masaryk University, Kotl\'{a}\v{r}sk\'{a} 2, 611 37
              Brno, Czech Republic\\
              \email{michalprisegen@gmail.com}
             }

   \date{Received Month day, 2023; accepted June day, 2023}

 
  \abstract
   {Open clusters (OCs) provide homogeneous samples of white dwarfs (WDs) with known distances, extinctions, and total ages. The unprecedented astrometric precision of \textit{\textit{Gaia}} allows us to identify many novel OC--WD pairs. Studying WDs in the context of their parent OCs makes it possible to determine the properties of WD progenitors and study the initial--final mass relation (IFMR).}
   {We seek to find potential new WD members of OCs in the solar vicinity. The analysis of OC members' parallaxes allows us to determine the OC distances to a high precision, which in turn enables us to calculate WD masses and cooling ages and to constrain the IFMR.}
   {We searched for new potential WD members of nearby OCs using the density-based machine learning clustering algorithm \texttt{HDBSCAN}. The clustering analysis was applied in five astrometric dimensions -- positions in the sky, proper motions and parallaxes, and in three dimensions where the positional information was not considered in the clustering analysis. The identified candidate OC WDs were further filtered using the photometric criteria and properties of their putative host OCs. The masses and cooling ages of the WDs were calculated via a photometric method using all available \textit{\textit{Gaia}}, Pan-STARRS, SDSS, and GALEX photometry. The WD progenitor masses were determined using the ages and metallicities of their host OCs.}
   {Altogether, 63 OC WD candidates were recovered, 27 of which are already known in the literature. We provide characterization for 36 novel WDs that have significant OC membership probabilities. Six of them fall into relatively unconstrained sections of the IFMR where the relation seems to exhibit nonlinear behavior. We were not able to identify any WDs originating from massive progenitors that would even remotely approach the widely adopted WD progenitor mass limit of 8~$\mathrm{M_{\odot}}$; this confirms the paucity of such objects residing in OCs and hints at a presence of velocity kicks for nascent WDs.}
   {}


   \maketitle
%

\section{Introduction}
White dwarfs (WDs) are compact objects that represent the ultimate stage of evolution of main sequence (MS) stars with zero-age MS masses less than about eight times that of the Sun \citep[e.g.,][]{weidemann_83}.\ This mass bracket encompasses the vast majority of all stars in the Galaxy. Due to their ubiquity, WDs comprise a significant component of Galactic populations. Since they no longer generate energy via central nuclear reactions, they gradually cool from their high initial temperatures in a predictable way. These properties make WDs a suitable tracer for studying Galactic history, the enrichment of the interstellar medium (ISM), the physics of matter under extreme conditions, and the progenitors of Type Ia supernovae \citep[SNe; see, e.g.,][for a general review]{althaus_2010,isern_2022,saumon_2022}. 

Due to their small sizes, WDs are intrinsically faint objects, with absolute magnitudes $\sim$10~mag fainter than MS stars of similar effective temperatures. Despite recent significant advances in astrometry, their faintness still poses problems for precise distance determination. Furthermore, due to their evolved nature, many of their properties, such as metallicity and mass, no longer reflect the properties of their progenitors. This is due to significant mass loss (especially in the last phases of stellar evolution) and the change in the observable chemical composition since elements heavier than H or He tend to diffuse downward into the WD interior due to the strong gravitational field \citep[e.g.,][]{schatzman_45,shapiro_83}. In the absence of any external accretion of matter, WDs relatively quickly end up with either a pure H (DA type) or He (DB type) atmosphere. As a consequence of these factors, observations of isolated WDs contain only limited information on the progenitor, and the limited precision of the measured distances also affects the derivation of other WD properties. However, if an association with another object or a group of stars with common properties can be established -- that is, when a WD is  part of a wide binary system with a stellar companion or a star cluster -- a number of these problems can be resolved and new research avenues are opened \citep[e.g.,][]{pasquini_2019,barrientos_2021}.

The most numerous class of star clusters are open clusters (OCs). They are gravitationally bound stellar aggregates (typically of $\sim 10^2$--$10^4$ stars) formed from a giant molecular cloud in a single star formation event. Open cluster member stars share several common properties, such as overall age, initial chemical composition (metallicity), overall kinematic properties in the context of the Galaxy, distance from the observer, and the amount of intervening ISM between OC stars and the observer (extinction or reddening; this may not hold for very young OCs). A large number of OC members and the use of robust statistics make the derivation of these parameters more accurate than what is otherwise typically achievable for an isolated star located in the Galactic field. The ages of known OCs cover a wide range, from a few million years to almost ten gigayears \citep[e.g.,][]{cantat-gaudin_2020, salaris_2004}. Metallicities span from subsolar [Fe/H]~=~-0.56 to super-solar values of about [Fe/H]~=~0.55 \citep{netopil_2016}. However, most OCs are relatively young and metal-rich, as most of the older OCs have long since dispersed into the Galactic field.

The fields of OC and WD research have been completely revolutionized with the advent of the \textit{\textit{Gaia}} satellite \citep{Gaia_summary}, especially its second data release \citep[GDR2;][]{Gaia_gdr2} and the early installment of its third data release \citep[EGDR3;][]{Gaia_edr3_summary}. EGDR3 presents accurate positions, proper motions, parallaxes, and broadband photometry in the $G$, $G_{\mathrm{BP}}$, and $G_{\mathrm{RP}}$ bands for almost 1.5 billion sources, with notable improvements over GDR2 in both the formal uncertainties and systematic effects \citep{Gaia_edr3_astrometry, Gaia_edr3_photometry}. Moreover, the number of sources has increased by 7\%, and there are proper motions and parallaxes available for 10\% more sources as compared to GDR2, which also yields  improved completenesses in dense areas of the sky \citep{Gaia_edr3_validation}. The complete third \textit{\textit{Gaia}} data release \citep[GDR3;][]{Gaia_dr3}, published in 2022, also includes source classifications \citep{Gaia_dr3_source_classification} and corrected photometry for some sources. The potential of a large-scale, astrometry-focused mission such as \textit{\textit{Gaia}} for WD science has been proven, with the known number of sources significantly expanded and a refinement of their physical properties \citep[e.g.,][]{jimenez-esteban_2018, bergeron_2019, fusillo_2019, kilic_2020, tremblay_2020, fusillo_2021, caron_2023}, enabling significant progress in this field. Similarly, the \textit{\textit{Gaia}} data have been instrumental in the discovery and characterization of many OCs \citep[e.g.,][and the references therein]{cantat-gaudin_2018, cantat-gaudin_2020, dias_2021}.

The fundamental properties of a WD -- its mass, cooling age (the time since the WD left the tip of the asymptotic giant branch), and internal and atmospheric composition -- can be obtained using spectroscopy, asteroseismology, or photometry. However, the spectroscopic and asteroseismic characterization of a large number of objects is observationally and computationally expensive. On the other hand, in the era of large, deep photometric all-sky surveys and with the availability of \textit{\textit{Gaia}} parallaxes and photometry, it is also possible to derive WD parameters using this type of data. However, photometry in the optical band by itself is unable to constrain the WD atmospheric composition, which is pivotal for the accurate determination of other WD physical parameters, as the atmosphere composition influences the way a WD cools. Despite this, it has been shown that the majority of observed WDs have DA-type atmospheres \citep[$\gtrsim 80$~\%; e.g.,][]{kepler_2016, kepler_2021}. This fraction appears to be even higher in OCs \citep[e.g.,][]{kalirai_2005,salaris_2019}. Masses and cooling ages for nearby WDs with negligible extinction can be determined using \textit{\textit{Gaia}} photometry and parallaxes to a precision of several percent \citep[e.g.,][]{bergeron_2019}.

Due to their nature, OCs provide homogeneous samples of WDs from coeval progenitors, all located at the same distance from the observer. Uncovering WD populations of OCs is pivotal for addressing numerous open issues in astrophysics. Of particular interest is the initial--final mass relation (IFMR), which links the star's initial mass, $M_{i}$, to its final mass, $M_{\mathrm{WD}}$, at the end of the stellar evolution, when the star has ultimately evolved into a WD. The IFMR constrains the amount of mass locked away in the WD and how much material is returned to the ISM. Also, the high-mass end of the IFMR marks the limit at which stars undergo core-collapse SNe. The accurate determination of the IFMR relies on obtaining a large and clean sample of OC WDs \citep[e.g.,][]{kalirai_2008,williams_2009,cummings_2018}.

 The IFMR determined in the literature contains extrinsic and intrinsic scatter. The intrinsic IFMR scatter is thought to be produced by the metallicity variation between the OCs that host WDs utilized for the IFMR determination because the stellar evolution, and especially the mass-loss in the last evolutionary stages, is thought to be significantly metallicity-dependent \citep[e.g.,][]{kalirai_2008,romero_2015,pastorelli_2019}. Another source of scatter may stem from the fact that the stellar evolution in the terminal phases might be inherently stochastic to an unknown degree and from the dispersion of the initial stellar rotational velocities. The main extrinsic components are the contamination from the WDs incorrectly assigned to OCs (i.e., physical nonmembers), measurement uncertainties and unaccounted systematic errors, incorrectly determined OC ages, and shortcomings in the stellar evolutionary models \citep[e.g.,][]{cummings_2018,cummings_2019}. 

A number of \textit{\textit{Gaia}}-based OC catalogs include tables of OC members stars, but the lower quality of astrometry and photometry at fainter magnitudes generally resulted in the exclusion of fainter stars from the analysis of OC parameters and the member tables in these catalogs. Because WDs tend to be located $\sim$10~mag below the OC MS in the OC color-magnitude diagram (CMD), only a small number of young WDs in the closest OCs are bright enough to be listed as cluster members in the current OC catalogs. 

Due to these issues and the availability of improved astrometry and photometry in GDR3, there is an opportunity to update the census of OC WDs. To obtain a larger sample of OC WDs, it is possible to cross-match the known and candidate cataloged WDs with an OC catalog using positional, parallax, and proper motion criteria, and to filter out the spurious WD--OC pairs using photometric constraints that are dependent on the distance, reddening, and age of the matched OC. Alternatively, it is also viable to extend the search of OC members to fainter magnitudes in order to reach the OC WD population.

In this paper we present an all-sky census of WDs in nearby Galactic OCs that aims to increase the number of known bona fide WD--OC pairs using data from the \textit{\textit{Gaia}} mission. The search for possible OC WD members was conducted using the unsupervised clustering algorithm HDBSCAN\footnote{\url{https://hdbscan.readthedocs.io}} \citep[Hierarchical Density-Based Spatial Clustering of Applications with Noise;][]{campello_2013,mcinnes_2017}.  The membership probabilities of the potential OC WDs are also estimated, allowing us to quantify our confidence in the physical association between the OC and the WD. Photometric data from \textit{\textit{Gaia}} were added to remove additional spurious pairings and, together with photometry from additional sources, used to calculate WD masses and cooling ages. These parameters, in conjunction with the properties of the host OC and stellar evolutionary models, were used to calculate the WD progenitor masses and examine the IFMR.

This paper is structured as follows. In Sect. 2 we describe the data used in this study and the workflow used to construct the preliminary list of OCs that host at least one WD candidate. In Sect. 3 we describe how the clustering analysis was conducted, comment on the data quality, re-derive the OC group parallaxes, collate other OC properties, and filter out the probably spurious WD-OC pairs. In Sect. 4 we calculate the WD masses, cooling ages, and masses of their progenitors and examine the IFMR. Finally, we provide a summary and conclusion in Sect. 5.

\section{Preliminary search}
Open clusters are observed throughout the Galactic disk, with the current census of well-established and characterized OCs numbering $\sim$1500  \citep[e.g.,][]{cantat-gaudin_2020,dias_2021}, some of the cataloged objects lying in the distance in excess of several kiloparsecs. Nevertheless, as the OCs are located predominantly near the Galactic Plane, the high source density, compounded with the high absorption due to the ISM in the Galactic disk, the OC census is thought to be incomplete beyond 1~kpc. This is evident by the high number of new OCs discovered using GDR2 and (E)GDR3 data \citep[e.g.,][]{castro-ginard_2018,castro-ginard_2021,liu_2019}. The visibility and the quality of astrometry and photometry of the intrinsically faint objects within the OCs, such as WDs, are especially affected by this and rapidly degrade with increasing distance. 

The limitations inherent to observing such faint objects are also clearly evident in the recent WD catalogs, such as that in \citet{fusillo_2021}, who present a collection of almost 1.3 million WD candidates based on EGDR3, they also list the probability of the objects being bona fide WDs (\texttt{Pwd}), which is computed using the spectroscopically confirmed WD sample from Sloan Digital Sky Survey (SDSS). Using the photogeometric distances, $d$, collated from \citet{bailer-jones_2021} of the likely WDs (\texttt{Pwd} > 0.5) in \citet{fusillo_2021}, it can be noted that 90~\% of the probable WDs lie at distances closer than 870~pc, with a median distance of $\sim$360~pc. However, there is a tail of WDs with $d>$1~kpc present in the sample of likely WDs.

According to WD evolutionary models \citep{bedard_2020}, a 10 Myr old, DA WD with $M\approx0.6 \, \mathrm{M_{\odot}}$ has intrinsic absolute $G_{\mathrm{BP}} \approx$ 9.7 mag. Assuming no extinction and \textit{\textit{Gaia}} $G_{\mathrm{BP}}$ mag limit of $\sim$20.0~mag, this translates into the distance limit of $\sim$1.1~kpc where reliable astrometry and photometry of WDs can be obtained considering \textit{\textit{Gaia}} capabilities in GDR3. This adopted limit is chosen due to the issues with \textit{\textit{Gaia}} $G_{\mathrm{BP}}$ photometry for very faint sources, which starts to be less reliable for sources fainter than $G_{\mathrm{BP}}\gtrsim$ 20~mag. Furthermore, the typical uncertainty of parallaxes and proper motions of WDs at this brightness is about 0.5~mas and 0.5~$\mathrm{mas} \, \mathrm{yr^{-1}}$, respectively, and rapidly deteriorating for objects fainter than 20~mag \citep{Gaia_edr3_summary}. Such uncertainties in the astrometry make it difficult to establish credible OC memberships for these objects beyond the distances of $\sim$1~kpc.

\citet{Gaia_edr3_photometry} suggest circumventing the problems with the $G_{\mathrm{BP}}$ photometry of faint objects by using $G - G_{\mathrm{RP}}$ as a color indicator; however, this approach is ill suited for WDs. The evolutionary tracks for WDs of various masses are more closely packed together in the $G - G_{\mathrm{RP}}$ versus $G$ CMD than in the more standard $G_{\mathrm{BP}}-G_{\mathrm{RP}}$ versus $G$ CMD, which makes the former CMD type less suitable for the WD characterization. Naturally, WDs younger than 10 Myr can be substantially brighter than 9.7 mag, making them detectable in the OCs further away than 1.1 kpc. However, the evolutionary tracks in the CMD are almost degenerate for such young WDs. Therefore, the analysis based on the CMDs constructed from the \textit{\textit{Gaia}} photometry is unable to provide accurate parameters for these objects. Therefore, considering the limits of \textit{\textit{Gaia}} photometry and astrometry, we adopt a distance limit of 1.1~kpc for OCs to be studied for the presence of WDs.

For the OC census, we considered the recent compilations by \citet{cantat-gaudin_2020} and \citet{dias_2021}, containing 2017 and 1743 OCs, respectively. We opted to adopt the richer catalog of \citet{cantat-gaudin_2020} for the preliminary WD search, discarding the well-known and well-studied OCs \object{Pleiades}, \object{Hyades,} and \object{Melotte 20}. Their proximity and large extent in the sky make them unsuited to be studied with the clustering technique used in this work. Moreover, the census of the single WD population of these OCs is most likely complete \citep[e.g.,][]{tremblay_2012,salaris_2018,heyl_22,miller_22}. Altogether, we retain 419 OCs within the distance limit of 1.1~kpc. These clusters span a large range of ages, from $\sim$6.5~Myr to $\sim$4.3~Gyr. 

To search for OCs hosting at least one isolated WD, we used the positions, radii, parallaxes, and proper motions reported in \citet{cantat-gaudin_2020} to query the \textit{\textit{Gaia}} archive for objects in the field of each OC as follows. First, a cone with 5$\times$\texttt{r50} radius around the center of each OC was chosen as the positional criterion. The parameter \texttt{r50} is used in \citet{cantat-gaudin_2020} as a measure of angular OC radius and is defined as a radius containing half of OC members. 
Second, using the mean values of OC parallax, proper motions, and their dispersions listed in \citet{cantat-gaudin_2020}, we applied cuts at 5$\sigma$ around these astrometric quantities to discard the objects from the cone search that are clearly not OC members.

Finally, only the stars with complete astrometric solutions (5p or 6p) were retained. We also applied the recommended basic quality cuts for the \textit{\textit{Gaia}} data, removing all sources with a renormalized unit weight error (\texttt{ruwe}) greater than 1.4, relative parallax error over 1.0, sources with no \textit{\textit{Gaia}} color, and very faint sources with $G_{\mathrm{BP}}>19.7$~mag. The magnitude limit was set to this value in order to accommodate the median OC extinction of $A_{G} \sim 0.4$~mag and to circumvent the problems in the BP photometry of faint blue sources as described in \citet{Gaia_edr3_photometry}.

Objects with \texttt{ruwe}$>$1.4 have high probabilities of having ill-behaved astrometric solutions yielding incorrect astrometric parameters for these objects \citep{Gaia_edr3_validation}. The inclusion of these objects in the analysis can lead to the detection of spurious OC members. Objects with \texttt{ruwe}$>$1.4 can occur in fields with high source densities, where close doubles that are not correctly handled in GDR3 can arise. High \texttt{ruwe} values can also indicate unresolved binarity. Therefore, filtering such objects also removes some contamination from unresolved binaries that are unsuitable for IFMR determination.

While the OC parameters given in \citet{cantat-gaudin_2020} are based on GDR2, the relative proximity of the studied OCs, together with the fact that the OC parameters are derived through robust statistics based on a large number of OC member stars, means that we do not expect the OC parameters to change appreciably in GDR3 and retaining the stars with the parallaxes and proper motions within 5$\sigma$ of the cataloged astrometric parameters of the studied OCs is sufficient to recover practically all physical  OC members. 

For the objects obtained in these preliminary queries within the OC fields, we corrected the reported GDR3 parallaxes using the zero-point correction described in \citet{Gaia_edr3_parallax_correction}. This correction depends on the type of the astrometric solution (5p or 6p solution), magnitude, color, and sky position\footnote{The zero-point is calculated using the python script provided in \url{https://gitlab.com/icc-ub/public/\textit{Gaia}dr3_zeropoint}.}.

For each cataloged OC, \citet{cantat-gaudin_2020} list the extinction values in $A_{V}$. We converted these to \textit{\textit{Gaia}} EDR3 passbands using the conversion factors in \citet{fusillo_2021}:
\begin{equation}
A_{G} = 0.835 A_{V},
\end{equation}
\begin{equation}
A_{BP} = 1.139 A_{V},
\end{equation}
\begin{equation}
A_{RP} = 0.650 A_{V}.
\end{equation}
Using these extinction factors and the cataloged parallaxes of each OC, the absolute de-reddened colors and magnitudes of all objects recovered by the queries have been derived, under the assumption of OC membership.

\begin{figure}
\centering
\includegraphics[width=\hsize]{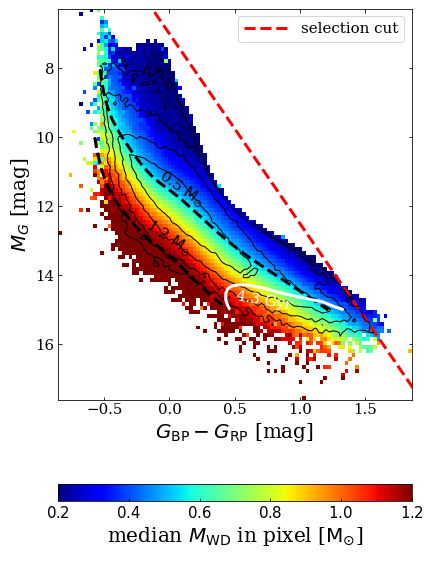}
\caption{Distribution of high-confidence WDs (sources with \texttt{Pwd}$>$0.5) with derived masses (assuming H atmosphere) from \citet{fusillo_2021} in the \textit{\textit{Gaia}} CMD. The dashed black lines are the cooling curves for 0.5 and 1.2 $M_{\odot}$ DA WDs from \citet{bedard_2020} and a 4.3~Gyr cooling age isochrone. The isolated OC WDs should all lie within the area bounded by these curves. The preliminary selection cut (dashed red line) should be flexible enough to detect these WDs even in the presence of significant uncertainties.}
\label{fig_1}
\end{figure}

To finally select all OCs potentially hosting at least one isolated WD, we constructed absolute color versus absolute magnitude $(G_{\mathrm{BP}}-G_{\mathrm{RP}})$ versus $M_\mathrm{G}$ CMD for all stars in the studied OC fields. We then applied a cut in the OC CMD below which all single WDs physically related to the OC are expected to be found:
\begin{equation}
\label{eqn:selection_criterion}
    M_\mathrm{G} > 7 + 5.5(G_{\mathrm{BP}}-G_{\mathrm{RP}}).
\end{equation}
The reason for the choice of the adopted criterion is illustrated in Fig.~\ref{fig_1}, which shows the CMD of high-confidence WDs (\texttt{Pwd}$>$0.5) from \citet{fusillo_2021}, color-coded according to their mass. WDs span a wide range of masses, from low-mass objects with $M_{\mathrm{WD}} \lesssim 0.5 \, M_{\odot}$ that have cores made up of He, up to high-mass WDs, which possibly host oxygen/neon cores, extending, in theory, all the way up to the Chandrasekhar limit. However, the lifetimes of the progenitors of He-core WDs are expected to be larger than the Hubble time, if isolated stellar evolution is assumed. Therefore, the majority of the observed population of these WDs is thought to be a consequence of close binary evolution involving mass transfer \citep[e.g.,][]{marsh_95, brown_2011, rebassa-mansergas_2011}. However, an alternative mechanism forming low-mass WDs from single metal-rich stars through extreme mass loss might be active as well \citep[e.g.,][]{hansen_2005, kilic_2007}. Excluding these objects significantly reduces the CMD parameter space where the single WDs physically associated with OCs can lie, and the cooling track for a 0.5~$M_{\odot}$ DA WDs approximately delineates the upper boundary of this parameter space. On the other hand, the lower boundary is delineated by the cooling curve of the most massive WDs.  In theory, the highest possible mass of a stable WD should be around $\sim$1.38~$\mathrm{M_{\odot}}$ \citep{nomoto_87}. However, so far, the most massive WDs found to be residing in OCs are only about 1.0~$\mathrm{M_{\odot}}$, with a handful of more massive objects ($\sim$1.2~$\mathrm{M_{\odot}}$) that were identified to be kicked out of their parent OCs \citep{richer_21,heyl_22,miller_22}. The paucity of the high-mass WDs within the OCs hints at the presence of a physical mechanism that imparts a velocity kick on the order of one to a few $\mathrm{km \, s^{-1}}$ to the nascent high-mass WDs \citep[e.g., ][]{feelhauer_2003,heyl_2007,davis_2008,fregeau_2009}. In light of this, the cooling curve for 1.2~$\mathrm{M_{\odot}}$ DA WD is plotted in Fig.~\ref{fig_1}. It is reasonable to expect that a vast majority of the WDs physically related to OCs will lie above this curve in the OC CMDs. Of course, the adopted selection criterion for the preliminary search is also sensitive to more massive WDs than this. Lastly, informed by the ages of the OCs listed in \citet{cantat-gaudin_2020}, we plot a 4.3~Gyr cooling isochrone contour in Fig.~\ref{fig_1}, which is the age of the oldest OC in the sample. Any WDs with cooling times higher than the age of the studied OC cannot have their origin within this particular OC.

While the adopted cut is very liberal and allows for significant uncertainties in the OC and candidate WD astrometry and photometry, it cannot be considered inclusive for all double WD systems and other WD binaries where the secondary significantly contributes to the total luminosity and the color of the system. However, since the IFMR only applies to isolated WDs, this is not detrimental to this particular scientific case.

After this first selection step, the list consisted of 238 OCs that potentially host at least one WD candidate that satisfies the adopted selection cut.

\section{Clustering analysis and establishing WD parameters}

In many cases, the preliminary selection of the OC stars in the previous section resulted in remarkably clean OC CMD diagrams, attesting to the quality of both the \textit{\textit{Gaia}} astrometry and the OC catalogs used in this work. However, for studying objects near the faint end of the brightness distribution, a more refined selection of OC stars is required. This is especially relevant for the fainter objects located below the OC MS, where the field star contamination that cannot be completely eliminated by the preliminary OC stars selection poses an issue.  

Since the primary aim of this study is to examine a semi-empirical IFMR that assumes that every analyzed WD is a product of a single-star evolution and is a coeval member of its associated OC, we need to search for WDs physically associated with OCs using the astrometric and photometric criteria. Firstly, in order for a WD to be considered a member of a particular OC, it needs to share a consistent position in the sky, distance, and proper motion with other members of the OC. Secondly, the bona fide isolated OC WDs occupy only the specific region of the OC CMD. This position is prescribed by the WD mass and its cooling age. Additionally, the cooling age of a WD associated with an OC cannot be greater than the age of the OC itself. This further constrains the possible positions that the OC WDs can occupy in the OC CMD.

The first criterion for finding WDs physically related to an OC is their astrometric membership. This implies that the WD and the rest of the OC stars are clustered together in the astrometric phase space, which can be ascertained using clustering analysis. In the following text, the term ``cluster'' does not refer to a physical OC but rather to a grouping of stars sharing similar properties in the N-dimensional astrometric parameter space.

A number of clustering methods have widely been utilized to search for structures in astronomical data with various degrees of success. However, for the detection of the structures such as the stellar streams, associations, and star clusters (including OCs) in the data set such as the one provided by the \textit{\textit{Gaia}} mission, only some of them are viable. Firstly, a suitable algorithm needs to be reasonably fast in order to conduct a clustering analysis on a large number of objects ($\sim 10^{6}$) in multiple dimensions within a reasonable time frame. Secondly, it must be able to discern between the clustered data and the data points belonging to the noise, as obviously, not all considered objects are necessarily part of a cluster within the data set. This is also grounded in reality, as only a small part of the stars in the Galaxy currently reside in OCs because OCs as gravitationally bound stellar groupings have relatively limited lifetimes. For a typical observed stellar field containing an OC, the majority of the stars will not belong to the OC but they will instead be part of the background/foreground Galactic disk population. Furthermore, the algorithm needs to be able to detect clusters of various shapes, sizes, and densities, as the properties of the astrometric phase space vary significantly depending on the position in the Galaxy. This also partly connects to the last requirement, where a suitable algorithm needs to ideally be provided as little information as possible prior to the analysis. Most importantly, it is generally not known a priori how many clusters are present in the studied data set.

The most prominent clustering methods used in the field of OC research are the UPMASK and pyUPMASK codes \citep{krone-martins_2014,pera_2021} and the DBSCAN \citep{ester_96} and HDBSCAN \citep{campello_2013,mcinnes_2017} algorithms. All of them have been widely utilized for OC searches and characterization as they satisfy most or all of the criteria outlined above \citep[e.g.,][]{castro-ginard_2018,cantat-gaudin_2018,castro-ginard_2019,cantat-gaudin_2019,castro-ginard_2020,hunt_2021,castro-ginard_2022,he_2022}.

In this work we utilize the HDBSCAN algorithm, which is intuitive to use and has proven to be very potent for studying OCs and other structures in the astrometric phase space \citep[e.g.,][]{kounkel_2019,tarricq_2022,moranta_2022,dungee_2022,chemel_2022}. HDBSCAN is a hierarchical clustering algorithm that builds on DBSCAN. The main advantages of using HDBSCAN over DBSCAN are that it is able to detect clusters of varying densities and that it does not require the non-intuitive epsilon hyperparameter that is required by DBSCAN, therefore providing results that are less biased than the DBSCAN output, which is significantly affected by the somewhat arbitrary choice of this clustering hyperparameter. Moreover, HDBSCAN is less sensitive to the selection of the clustering parameters than HDBSCAN while also being slightly faster. 

The main required hyperparameter (i.e., parameter specified by the user) that controls the performance of HDBSCAN is \texttt{min\_cluster\_size}, which sets the minimum number of data points required to form a cluster. There is a possibility to specify a second hyperparameter, \texttt{min\_samples}, that determines how conservative the clustering is, with larger values of this parameter yielding more points classified as noise and clusters confined into progressively denser regions of the phase space. By default, \texttt{min\_samples} is set to the same value as \texttt{min\_cluster\_size}.

HDBSCAN also offers a choice of two possible clustering approaches that dictate how the algorithm selects clusters from the cluster tree hierarchy. The default method is `excess of mass,'' which tends to select about one or two populous clusters and a number of smaller clusters. Another option is to use the ``leaf'' method, which yields a larger number of more homogeneous clusters. \citet{hunt_2021} note that the leaf clustering method generally performs better in the identification of OCs. Therefore, we adopted the leaf approach in this work.

\subsection{HDBSCAN clustering}

A new search was conducted using the coordinates of the OCs preselected in Sect. 2. based on their astrometric parameters listed in \citet{cantat-gaudin_2020}. The query criteria are similar to the ones in Sect. 2, albeit with some notable differences.\ First, we used a cone search of 6$\times$\texttt{r\_50} instead of 5$\times$\texttt{r\_50}. This was done to increase the number of field stars in the OC surroundings to increase the contrast of the OC in the positional space for the clustering algorithm.

Similarly, the studied proper motion parameter space was also increased to $\pm$10~$\mathrm{mas \, yr^{-1}}$ around the central values of the OC proper motion. This is sufficient since the maximum standard deviation of the proper motion for the preselected OCs is only about $\sim$1.3~$\mathrm{mas \, yr^{-1}}$. Finally, the parallax was limited to objects with $\varpi>0.75$~mas in order to reduce to crowding of the parameter space by background objects. The five astrometric parameters used for the clustering analysis ($\alpha, \delta, \mu_{\alpha^{\star}}, \mu_{\delta}, \varpi$) were rescaled using \texttt{RobustScaler} from \texttt{scikit-learn} \citep{scikit-learn} to have a zero median and a unit interquartile range. 

A typical OC comprises a dense compact core and a more extended sparse halo/corona, sometimes with tidal tails. Both these extended structures can span several tens of pc from the OC center \citep[e.g.,][and the references therein]{nilakshi_2002,meingast_2021,tarricq_2022,bhatta_2022}. Therefore, the positional constraints are not as informative for clustering analysis as the other dimensions, such as proper motions. This could be resolved if it was possible to apply different weights to the dimensions used for clustering, but HDBSCAN does not offer this feature. Because of this, we ran HDBSCAN in two ways -- in 5D ($\alpha, \delta, \mu_{\alpha^{\star}}, \mu_{\delta}, \varpi$) and in 3D ($\mu_{\alpha^{\star}}, \mu_{\delta}, \varpi$). This was done because including the positional criteria in the clustering analysis may exclude physical members in the OC outskirts. On the other hand, excluding the positional criteria may introduce more contamination in the recovered clusters. For both the 5D and 3D analysis, we tested the clustering performance with different input hyperparameter combinations on a representative subset of 20 OCs that were picked from the studied OCs. This was necessary because, for some hyperparameter combinations, a significant portion of the studied OCs was not detected by the algorithm in their fields and was instead assigned to the noise. Also, for some combinations, only the dense OC cores were detected. Therefore, most importantly, we aimed to maximize the number of the detected OCs. Of secondary importance was the completeness of the OC member population, which was assessed by the comparison with the OC members listed in \citet{cantat-gaudin_2020}. Out of the tested combinations, \texttt{min\_cluster\_size=30} \& \texttt{min\_samples=3} for the 5D case, and \texttt{min\_cluster\_size=20} \& \texttt{min\_samples=2} for the 3D case were found to offer the best performance and were adopted in the further analysis.

Depending on the input parameters, HDBSCAN typically finds several tens of clusters in each of the studied OC fields. To facilitate easier matching of the physical OC of interest to the statistical clusters detected by HDBSCAN, we computed the median proper motion values for each detected cluster and retained only those with the median proper motion within 3$\sigma$ of the value listed in \citet{cantat-gaudin_2020} for the targeted OC. Doing so significantly reduced the number of clusters that needed to be inspected. The cluster corresponding to the physical OC was determined by plotting the CMD, parallax histogram, vector-point, and position diagram for all clusters and overlaying the data for the OC members cataloged in \citet{cantat-gaudin_2020} on top of these diagrams and choosing the cluster that offered the best match.

\subsection{Membership probability}

Aside from its inability to apply different weights to different clustering dimensions, another downside of HDBSCAN is that it does not take into consideration uncertainties in the input data. To rectify this and to compute membership probabilities of OC WD candidates we used a Monte Carlo approach, taking the uncertainties and correlations in the OC WD candidate astrometry into consideration. We conducted 100 runs, each time drawing a new set of parallax and proper motion values for the WD candidate based on a multivariate normal distribution, which was constructed using the OC WD candidate astrometric quantities, their uncertainties, and the correlation coefficients listed in the GDR3 catalog. Having obtained these new values for the considered object, we reapplied the HDBSCAN clustering on the OC field 100 times. The probability of the WD candidate membership in the particular OC is then the ratio of the number of outcomes where the HDBSCAN assigns the WD candidate to the OC and the number of outcomes when the WD is assigned elsewhere -- either to the field population or to some other statistical or physical cluster present in the OC field. For instance, a WD assigned to the OC in 50 runs of the algorithm is assigned a membership probability of 0.5.

\subsection{Data quality}
The faintness of our objects of interest and their position within or in projection to OCs located in the Galactic plane makes it important to assess the quality of the \textit{\textit{Gaia}} astrometry and photometry of these objects, which is the base of the further analysis. Some preliminary filtering using data quality indicators and source characteristics available from the \textit{\textit{Gaia}} catalog (cuts based on \texttt{ruwe}, relative parallax error, and brightness) has already been done in the OC field queries in Sect. 2. Such filtering may not be sufficient to remove all unreliable sources. However, other quality indicators can be used to assess the quality and reliability of the \textit{\textit{Gaia}} data.

The candidate OC WDs were cross-matched with the catalog of \citet{rybizky_2022} who provide the astrometric \texttt{fidelity} flag. This is a reliability diagnostic that is based on a neural network classifier that was trained on the GDR3 astrometric entries from a set of presumably bad and presumably trustworthy GDR3 sources. The value of the \texttt{fidelity} flag varies between 0 and 1.0, where 1.0 indicates a perfectly reliable solution, whereas 0.0 indicates a source with untrustworthy astrometry. Objects with spurious astrometric solutions can be filtered out from the analysis by removing the sources with the \texttt{fidelity} flag lower than 0.5.

It is also possible to identify the sources with unreliable photometry. This can be done using the corrected value of the flux excess factor \citep[\texttt{phot\_bp\_rp\_excess\_factor} in the \textit{\textit{Gaia}} archive, ][]{Gaia_edr3_summary, Gaia_edr3_photometry}, $C^{\star}$. For well-behaved point sources, $C^{\star}$ should have a flat distribution when plotted with \textit{\textit{Gaia}} color that is centered on zero. Significant deviations from this trend may indicate that the source photometry may be contaminated by flux from objects in its proximity. When culling the sources with potentially affected photometry, we discarded objects with $|{C^{\star}}|/\sigma_{C^{\star}} (G)$~$>$~5, where $\sigma_{C^{\star}}$ represents the 1$\sigma$ scatter expected to be present for well-behaved sources, which is computed as a function of $G$ according to Eq. 18 of \citet{Gaia_edr3_photometry}.

\subsection{OC parallaxes}
 We computed group parallaxes of the recovered OCs using the member stars recovered by HDBSCAN and after applying a 2-$\sigma$ clipping around the median parallax value of the OC members and following the procedure described in \citet{maiz_2021}. Due to a typically large number of the recovered OC member stars, the statistical uncertainty of the group OC parallax is very small, only up to a few $\mu \mathrm{as}$. However, the \textit{Gaia} parallaxes have an angular covariance that places limits on the minimum achievable uncertainty for the OC group parallaxes, and the errors stemming from these covariances dominate the total error budget. We list the derived group parallaxes of the OCs that are potential hosts of the newly characterized WDs in Table~\ref{table_oc}.

\subsection{Supplementing the WD sample with the Hunt \& Reffert catalog}
Recently\footnote{This catalog was released as a preprint during the first round of revisions of this paper.}, \citet{hunt_2023} constructed a large catalog of star clusters, also using a methodology based on HDBSCAN. Altogether, they list 7167 clusters, with the majority of them being OCs. For each cluster, they also provide a list of members with membership probabilities. Aside from 739 newly detected clusters, the novelty of this work is the depth of their list of cluster members, where the adoption of the \texttt{fidelity} criterion \citep{rybizky_2022} allows the inclusion of the objects up to $G \sim $~20~mag for some OCs, which is notably deeper than the membership lists constructed using the simple magnitude cut \citep[adopted in, e.g.,][]{cantat-gaudin_2018, cantat-gaudin_2020}. This means that their OC members lists are more likely to contain a number of previously unstudied WDs than the previous works in the literature.

In order to leverage this catalog, we considered a sample of astrometrically and photometrically reliable \citep[as defined in][]{hunt_2023} OCs with distances below 2~kpc, which yielded 2257 OCs. Taking into consideration the cluster members with membership probabilities higher than 0.5, we calculate their absolute magnitudes and de-reddened colors using the OC parameters as listed in \citet{hunt_2023} and apply the criterion from Eq.~\ref{eqn:selection_criterion} to identify the possible OC hosting WDs. This first selection identified, after excluding the objects with \texttt{ruwe}$>$1.4 and $|{C^{\star}}|/\sigma_{C^{\star}} (G)$~$>$~5, 67 OCs potentially hosting at least one WD in this catalog. For potential OCs hosting WDs, we also calculated the group parallaxes and their errors using the methodology outlined in Sect.~3.4.

\subsection{Other OC parameters}
We make use of the published OC catalogs to get other OC physical parameters, such as extinction, total age, and metallicity. \citet{cantat-gaudin_2020} do not list OC metallicities and for OC extinctions and total ages, they do not provide explicit uncertainty values in their data table. However, they report that the OC extinction uncertainties typically span the range 0.1--0.2~mag, and for the total ages ($\log t_{\mathrm{OC}}$) the uncertainty ranges 0.15--0.25 for young OCs and 0.1--0.2 for the older objects. \citet{dias_2021} provide explicit uncertainties for OC extinctions, total ages, and metallicities. However, they do not list the parameters for some of the ``UBC'' and ``UPK'' OCs \citep{sim_2019,castro-ginard_2020} that have been identified as likely WD hosts. Therefore, we adopt the OC parameters from \citet{dias_2021}, except for the few OCs not included in their catalog. For these objects, we adopt the parameters listed in \citet{cantat-gaudin_2020}, with the conservative uncertainty estimates of 0.2~mag for the extinction, $\log t_{\mathrm{OC}}$/yr of 0.2 for the total OC age, and we assume the OC metallicities to be solar. If the OC is not present in both of these catalogs, we adopt its age and extinction from \cite{hunt_2023}, where we again assume solar metallicities, as this catalog also does not contain this information. The OC parameters collated from \citet{dias_2021}, \citet{cantat-gaudin_2020}, and \citet{hunt_2023} are also tabulated in Table~\ref{table_oc}.

\subsection{Position of the WDs in the OC CMDs and preliminary cooling ages}

\begin{figure}
\centering
\includegraphics[width=\hsize]{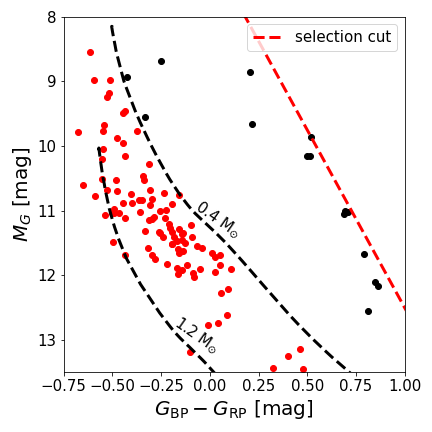}
\caption{\textit{\textit{Gaia}} CMD of WD candidates with probabilities of OC membership from the 5D, 3D, or \citet{hunt_2023} analysis $\geq$ 0.5. The objects selected for the subsequent analysis are marked in red. The error bars are omitted for clarity. The dashed black lines represent the cooling curves for 0.4 and 1.2 $M_{\odot}$ DA WDs from \citet{bedard_2020}, and the dashed red line indicates the preselection cut employed in Sect. 2. The 4.3~Gyr cooling age isochrone seen in Fig.~\ref{fig_1} is outside the bounds of the figure. Discarded objects (in black) lying outside of the region delineated by the 0.4~$M_{\odot}$ cooling curve and the 4.3~Gyr cooling age isochrone are not considered in the subsequent analysis.}
\label{fig_2}
\end{figure}

Figure~\ref{fig_2} shows the CMD of the OC WD sample in \textit{\textit{Gaia}} filters with the cooling tracks for 0.4 and 1.2~$\mathrm{M_{\odot}}$ DA WDs from \citet{bedard_2020}. We used this CMD and the 0.4~$\mathrm{M_{\odot}}$ cooling track to select the WD candidates that are consistent with being isolated stars. No objects needed to be cut based on the excessive cooling age as compared to the maximum total age of the studied OCs.

Additional filtering can be done using by comparing the WD cooling age ($t_{\mathrm{cool}}$) with the overall OC age ($t_{\mathrm{OC}}$), as WDs that are OC members cannot have $t_{\mathrm{cool}}$ larger than $t_{\mathrm{cool}}$. The coordinates of WD candidates in the OC CMD constructed from the \textit{\textit{Gaia}} photometry can be converted into preliminary $t_{\mathrm{cool}}$ estimates by the interpolation between the set of cooling tracks for DA WDs with C/O core \citep{bedard_2020}. 

To account for the uncertainties in WD candidate photometry, OC group parallax, and OC extinction, we performed $10^4$ Monte Carlo draws, each time drawing values of these parameters from normal distributions (assumed to be independent of each other) characterized by their mean values and 1$\sigma$ errors. These were used to calculate the absolute $G$ magnitude and de-reddened color of an OC WD, which were in turn used to interpolate preliminary $t_{\mathrm{cool}}$ from the WD cooling tracks for each simulation run using the Python tool of \citet{cheng_code}. Here and in the subsequent analysis, the listed values are medians of the values obtained from the simulations and the quoted uncertainties were calculated using the $16^{\mathrm{th}}$ and $84^{\mathrm{th}}$ percentiles of the resulting distributions. After this step, we excluded the objects with their median value of $t_{\mathrm{cool}}$ estimate higher than the $t_{\mathrm{OC}}$ of the associated OC, yielding the final sample of 77 possible OC WDs, listed in Table~\ref{wds_all}.

\subsection{Issue of contamination and uncertainties in the OC parameters}
It needs to be noted that constructing a clean sample of OC WDs and conducting meaningful studies that build on this sample are extremely challenging due to the low brightness of these objects and, despite a lot of recent progress, also due to significant uncertainties in the OC parameters listed in the recent catalogs. The issue is apparent also in the Fig.~\ref{fig_2}, where it can be seen that some objects, despite their high OC astrometric membership probability, are located blueward of the limit where the WDs should lie in the CMD. This can be attributed to various reasons, such as uncorrected issues with the photometry, spurious matching of the WD to the OC, or an overestimated extinction of the OC.

\setcounter{table}{1}
\begin{table*}
\caption{Literature WDs recovered as OC members in the clustering analysis. WD masses ($M_{\mathrm{WD}}$) 
and progenitor masses ($M_{\mathrm{i}}$) have been adopted from the listed reference.}          
\label{wds_literature_parameters}      
\centering          
\begin{tabular}{llllll}     
\hline\hline  
\textit{\textit{Gaia}} DR3 \texttt{source\_id} & associated OC & $M_{\mathrm{WD}}$ & $M_{\mathrm{i}}$ & reference \\
 &  & ($\mathrm{M_{\odot}}$) & ($\mathrm{M_{\odot}}$) & \\ 
\hline
4008511467191955840 & Melotte 111  & ${0.90}_{{-0.04}}^{+{0.04}}$ & ${4.77}_{{-0.97}}^{+{5.37}}$ & \citet{dobbie_2009} \\
5290767695648992128 & NGC 2516  & ${0.92}_{{-0.03}}^{+{0.03}}$ & ${4.62}_{{-0.09}}^{+{0.11}}$ & \citet{cummings_2018} \\
5290834387897642624 & NGC 2516  & ${0.97}_{{-0.03}}^{+{0.03}}$ & ${4.98}_{{-0.12}}^{+{0.15}}$ & \citet{cummings_2018} \\
5290719287073728128 & NGC 2516  & ${0.92}_{{-0.03}}^{+{0.03}}$ & ${4.89}_{{-0.11}}^{+{0.12}}$ & \citet{cummings_2018} \\
5290720695823013376 & NGC 2516 & ${0.98}_{{-0.04}}^{+{0.04}}$ & ${4.83}_{{-0.09}}^{+{0.11}}$ & \citet{cummings_2018} \\
661297901272035456 & NGC 2632 & ${0.757}_{{-0.025}}^{+{0.025}}$ & ${2.96}_{{-0.03}}^{+{0.03}}$ & \citet{cummings_2018} \\ 
661311267210542080 & NGC 2632 & ${0.830}_{{-0.028}}^{+{0.028}}$ & ${3.36}_{{-0.10}}^{+{0.12}}$ & \citet{cummings_2018}  \\ 
661270898815358720 & NGC 2632 & ${0.87}_{{-0.04}}^{+{0.04}}$ & ${2.78}_{{-0.20}}^{+{0.17}}$ & \citet{salaris_2019} \\
661010005319096192 & NGC 2632 & ${0.76}_{{-0.02}}^{+{0.02}}$ & ${3.05}_{{-0.29}}^{+{0.27}}$ & \citet{salaris_2019} \\
661353224747229184 & NGC 2632 & ${0.81}_{{-0.02}}^{+{0.02}}$ & ${3.02}_{{-0.28}}^{+{0.25}}$ & \citet{salaris_2019} \\
659494049367276544 & NGC 2632 & ${0.81}_{{-0.04}}^{+{0.04}}$ & ${3.16}_{{-0.32}}^{+{0.31}}$ & \citet{salaris_2019} \\
660178942032517760 & NGC 2632 & ${0.77}_{{-0.03}}^{+{0.03}}$ & ${3.28}_{{-0.37}}^{+{0.37}}$ & \citet{salaris_2019} \\
665139697978259200 & NGC 2632 & ${0.69}_{{-0.05}}^{+{0.05}}$ & ${3.01}_{{-0.28}}^{+{0.26}}$ & \citet{salaris_2019} \\
662798086105290112 & NGC 2632 & ${0.73}_{{-0.02}}^{+{0.02}}$ & ${2.92}_{{-0.24}}^{+{0.22}}$ & \citet{salaris_2019} \\
662998983199228032 & NGC 2632 & ${0.80}_{{-0.04}}^{+{0.04}}$ & ${3.25}_{{-0.28}}^{+{0.28}}$ & \citet{salaris_2019} \\
664325543977630464 & NGC 2632 & ${0.75}_{{-0.04}}^{+{0.04}}$ & ${3.02}_{{-0.28}}^{+{0.26}}$ & \citet{salaris_2019} \\
661841163095377024 & NGC 2632 & ${0.78}_{{-0.02}}^{+{0.02}}$ & ${3.50}_{{-0.46}}^{+{0.51}}$ & \citet{salaris_2019} \\
5340219811654824448 & NGC 3532 & ${0.75}_{{-0.03}}^{+{0.03}}$ & ${3.36}_{{-0.01}}^{+{0.01}}$ & \citet{cummings_2018} \\
2170776080281869056 & NGC 7092 & ${0.95}_{{-0.02}}^{+{0.02}}$ & ${5.40}_{{-0.60}}^{+{0.60}}$ & \citet{richer_21} \\
4088108859141437056 & Ruprecht 147 & ${0.66}_{{-0.02}}^{+{0.02}}$ & ${1.68}_{{-0.01}}^{+{0.01}}$ & \citet{marigo_2020} \\
4087806832745520128 & Ruprecht 147 & ${0.67}_{{-0.02}}^{+{0.02}}$ & ${1.66}_{{-0.01}}^{+{0.01}}$ & \citet{marigo_2020} \\
4183937688413579648 & Ruprecht 147 & ${0.67}_{{-0.02}}^{+{0.02}}$ & ${1.68}_{{-0.01}}^{+{0.01}}$ & \citet{marigo_2020} \\
4183978061110910592 & Ruprecht 147 & ${0.69}_{{-0.02}}^{+{0.02}}$ & ${1.73}_{{-0.01}}^{+{0.01}}$ & \citet{marigo_2020} \\
4184169822810795648 & Ruprecht 147 & ${0.68}_{{-0.02}}^{+{0.02}}$ & ${1.66}_{{-0.01}}^{+{0.01}}$ & \citet{marigo_2020} \\
4183919237232621056 & Ruprecht 147 & ${0.67}_{{-0.02}}^{+{0.02}}$ & ${1.65}_{{-0.01}}^{+{0.01}}$ & \citet{marigo_2020} \\
506862078583709056 & Stock 2 & ${0.99}_{{-0.03}}^{+{0.03}}$ & ${5.50}_{{-1.50}}^{+{1.50}}$ & \citet{richer_21} \\
1992469104239732096 & Stock 12 & ${0.94}_{{-0.02}}^{+{0.02}}$ & ${3.70}_{{-0.40}}^{+{0.40}}$ & \citet{richer_21} \\
\hline
\end{tabular}
\end{table*}

\begin{figure*}
\centering
\includegraphics[width=\hsize]{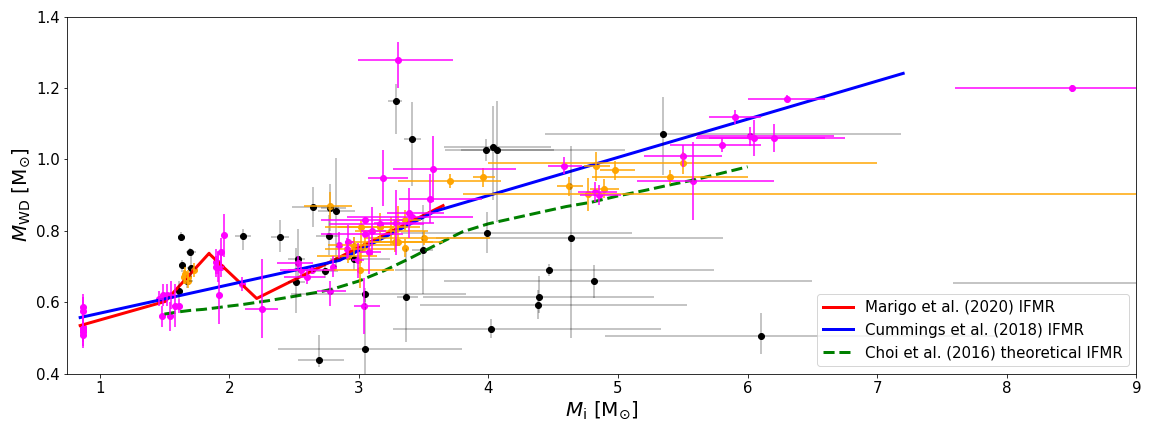}
\caption{IFMR for WDs within star clusters. Black points represent the newly characterized WDs that have been 
recovered to be OC members in this work, tabulated in Table~\ref{wds_novel}. The orange points denote known WDs collated from the literature (see Table~\ref{wds_literature_parameters}) that have also been 
recovered in the clustering analysis. Their masses ($M_{\mathrm{WD}}$) and the progenitor masses ($M_{\mathrm{WD}}$) are adopted
from the sources listed in Table~\ref{wds_literature_parameters}. In purple, we plot WDs that are hosted by star clusters not considered in this study (due to the adopted distance limit or spatial constraints) listed in \citet{salaris_2018}, \citet{cummings_2018}, \citet{richer_2019}, \citet{marigo_2020}, \citet{canton_2021}, \citet{heyl_22}, and \citet{miller_22}, and central stars of planetary nebulae are from \citet{fragkou_2019,fragkou_2022}. Several IFMR fits are also shown: semi-empirical fits from \citet{marigo_2020} (in red) and \citet{cummings_2018} (in cyan) and the theoretical fit from \citet{choi_2016} (the dashed green line).}
\label{fig_ifmr}
\end{figure*}

\section{WD properties and the IFMR}
There is a large volume of literature dealing with finding OC WDs, deriving their properties, and using them to study the IFMR. The analysis conducted in the previous sections allowed us to recover and independently confirm the membership of several OC WDs previously studied in the literature. We collate them and their derived properties relevant to the IFMR in Table~\ref{wds_literature_parameters}. Since they were predominantly subjects of dedicated studies, often also employing spectroscopic data, we do not re-derive their parameters again in this work.

\begin{table*}
\caption{Novel or previously uncharacterized WDs recovered as OC members in the clustering analysis. }          
\label{wds_novel}      
\centering          
\begin{tabular}{lllllll}     
\hline\hline  
\textit{\textit{Gaia}} DR3 \texttt{source\_id} & associated OC & $P_{\mathrm{max}}$ & avail. photometry & $M_{\mathrm{WD}}$ & $t_{\mathrm{cool}}$ & $M_{\mathrm{i}}$ \\
 &  &  & ($\mathrm{M_{\odot}}$) & (Myr) & ($\mathrm{M_{\odot}}$) \\ 
\hline
2879428195013510144 & Alessi 22 & 1.0 & \textit{Gaia}, PS, \textit{GALEX} & ${0.62}_{{-0.02}}^{+{0.02}}$ & ${207}_{{-8}}^{+{7}}$ & ${3.05}_{{-0.57}}^{+{0.78}}$ \\
4519349757798439936 & Alessi 62 & 0.8 & \textit{Gaia}, PS, SDSS, \textit{GALEX} & ${0.78}_{{-0.04}}^{+{0.05}}$ & ${1}_{{0}}^{+{0}}$ & ${2.39}_{{-0.07}}^{+{0.06}}$ \\
391939027303287040 & Alessi 94 & 1.0 & \textit{Gaia}, PS & ${0.62}_{{-0.05}}^{+{0.06}}$ & ${11}_{{-2}}^{+{2}}$ & ${4.39}_{{-0.89}}^{+{0.89}}$ \\
386116254240723712 & Alessi 94 & 1.0 & \textit{Gaia}, PS & ${0.59}_{{-0.04}}^{+{0.04}}$ & ${15}_{{-2}}^{+{3}}$ & ${4.38}_{{-0.92}}^{+{1.15}}$ \\
5798954758752816512 & CWNU 41 & 1.0 & \textit{Gaia} & ${0.47}_{{-0.11}}^{+{0.15}}$ & ${92}_{{-30}}^{+{40}}$ & ${3.05}_{{-0.67}}^{+{0.75}}$ \\
335525529520188032 & HSC 1155 & 1.0 & \textit{Gaia}, PS & ${0.53}_{{-0.03}}^{+{0.03}}$ & ${8}_{{-1}}^{+{1}}$ & ${4.02}_{{-0.76}}^{+{1.31}}$ \\
5257090272981848192 & HSC 2304 & 0.9 & \textit{Gaia} & ${0.78}_{{-0.28}}^{+{0.26}}$ & ${7}_{{-5}}^{+{5}}$ & ${4.64}_{{-1.04}}^{+{1.17}}$ \\
4502736137087916032 & HSC 381 & 0.9 & \textit{Gaia}, PS, \textit{GALEX} & ${0.65}_{{-0.01}}^{+{0.01}}$ & ${3}_{{0}}^{+{0}}$ & ${9.41}_{{-1.83}}^{+{4.91}}$ \\
6653882460188145152 & Mamajek 4 & 0.7 & \textit{Gaia}, \textit{GALEX} & ${0.69}_{{-0.02}}^{+{0.02}}$ & ${65}_{{-2}}^{+{3}}$ & ${2.74}_{{-0.12}}^{+{0.10}}$ \\
4662157454731788672 & NGC 1901 & 0.6 & \textit{Gaia} & ${0.66}_{{-0.09}}^{+{0.09}}$ & ${64}_{{-10}}^{+{14}}$ & ${2.52}_{{-0.07}}^{+{0.06}}$ \\
4659513404187412736 & NGC 1901 & 0.8 & \textit{Gaia} & ${0.72}_{{-0.07}}^{+{0.08}}$ & ${64}_{{-7}}^{+{8}}$ & ${2.53}_{{-0.07}}^{+{0.06}}$ \\
2931807898171448320 & NGC 2358 & 0.8 & \textit{Gaia}, PS & ${0.86}_{{-0.16}}^{+{0.15}}$ & ${2}_{{-1}}^{+{3}}$ & ${2.83}_{{-0.14}}^{+{0.14}}$ \\
5289447182180342016\tablefootmark{a} & NGC 2516 & 0.9 & \textit{Gaia}, \textit{GALEX} & ${1.03}_{{-0.03}}^{+{0.03}}$ & ${67}_{{-3}}^{+{2}}$ & ${3.98}_{{-0.32}}^{+{0.52}}$ \\
5294015515555860608\tablefootmark{a} & NGC 2516 & 1.0 & \textit{Gaia} & ${1.03}_{{-0.15}}^{+{0.12}}$ & ${63}_{{-9}}^{+{7}}$ & ${4.03}_{{-0.37}}^{+{0.45}}$ \\
5338675689360848256 & NGC 3532 & 0.6 & \textit{Gaia} & ${1.16}_{{-0.09}}^{+{0.05}}$ & ${37}_{{-21}}^{+{7}}$ & ${3.28}_{{-0.06}}^{+{0.05}}$ \\
5337742307052922752 & NGC 3532 & 0.8 & \textit{Gaia} & ${0.75}_{{-0.12}}^{+{0.13}}$ & ${94}_{{-16}}^{+{19}}$ & ${3.49}_{{-0.08}}^{+{0.08}}$ \\
5339898655484772864 & NGC 3532 & 0.8 & \textit{Gaia} & ${0.62}_{{-0.13}}^{+{0.14}}$ & ${59}_{{-17}}^{+{25}}$ & ${3.36}_{{-0.07}}^{+{0.09}}$ \\
5340530320614001792 & NGC 3532 & 0.5 & \textit{Gaia} & ${1.06}_{{-0.13}}^{+{0.10}}$ & ${75}_{{-13}}^{+{11}}$ & ${3.41}_{{-0.06}}^{+{0.07}}$ \\
5340165355769599744 & NGC 3532 & 0.6 & \textit{Gaia} & ${1.03}_{{-0.20}}^{+{0.14}}$ & ${211}_{{-48}}^{+{102}}$ & ${4.07}_{{-0.28}}^{+{0.99}}$ \\
4477168746525464064\tablefootmark{b} & NGC 6633 & 0.7 & \textit{Gaia}, PS & ${0.86}_{{-0.06}}^{+{0.07}}$ & ${40}_{{-2}}^{+{2}}$ & ${2.78}_{{-0.06}}^{+{0.12}}$ \\
4477214475044842368\tablefootmark{c} & NGC 6633 & 0.6 & \textit{Gaia}, PS & ${0.79}_{{-0.06}}^{+{0.06}}$ & ${11}_{{-2}}^{+{1}}$ & ${2.77}_{{-0.10}}^{+{0.08}}$ \\
342523646152426368 & NGC 752 & 0.7 & \textit{Gaia}, PS, \textit{GALEX} & ${0.79}_{{-0.04}}^{+{0.02}}$ & ${171}_{{-5}}^{+{5}}$ & ${2.10}_{{-0.06}}^{+{0.06}}$ \\
4087833117945955840 & Ruprecht 147 & 0.7 & \textit{Gaia}, PS, \textit{GALEX} & ${0.71}_{{-0.02}}^{+{0.01}}$ & ${215}_{{-5}}^{+{5}}$ & ${1.64}_{{-0.02}}^{+{0.02}}$ \\
4183928888026931328 & Ruprecht 147 & 0.9 & \textit{Gaia}, PS & ${0.63}_{{-0.02}}^{+{0.03}}$ & ${139}_{{-7}}^{+{6}}$ & ${1.61}_{{-0.02}}^{+{0.03}}$ \\
4183847562828165248 & Ruprecht 147 & 1.0 & \textit{Gaia}, PS & ${0.66}_{{-0.02}}^{+{0.01}}$ & ${433}_{{-26}}^{+{13}}$ & ${1.67}_{{-0.03}}^{+{0.04}}$ \\
4184148073089506304 & Ruprecht 147 & 0.8 & \textit{Gaia}, PS & ${0.74}_{{-0.01}}^{+{0.01}}$ & ${501}_{{-38}}^{+{43}}$ & ${1.70}_{{-0.04}}^{+{0.04}}$ \\
4183926006112672768\tablefootmark{d} & Ruprecht 147 & 0.9 & \textit{Gaia}, PS & ${0.70}_{{-0.02}}^{+{0.05}}$ & ${542}_{{-54}}^{+{9}}$ & ${1.71}_{{-0.03}}^{+{0.04}}$ \\
4184196073644880000 & Ruprecht 147 & 0.8 & \textit{Gaia}, PS & ${0.78}_{{-0.01}}^{+{0.01}}$ & ${255}_{{-5}}^{+{6}}$ & ${1.63}_{{-0.02}}^{+{0.03}}$ \\
507105143670906624 & Stock 2 & 0.5 & \textit{Gaia}, PS & ${0.72}_{{-0.03}}^{+{0.06}}$ & ${312}_{{-38}}^{+{18}}$ & ${2.96}_{{-0.18}}^{+{0.28}}$ \\
507555904779576064 & Stock 2 & 0.9 & \textit{Gaia}, PS & ${0.44}_{{-0.02}}^{+{0.07}}$ & ${122}_{{-22}}^{+{5}}$ & ${2.69}_{{-0.16}}^{+{0.19}}$ \\
506848643933335296 & Stock 2 & 0.6 & \textit{Gaia}, PS & ${0.87}_{{-0.05}}^{+{0.06}}$ & ${89}_{{-7}}^{+{9}}$ & ${2.65}_{{-0.16}}^{+{0.20}}$ \\
2174431990805230208 & Theia 248 & 1.0 & \textit{Gaia}, PS & ${1.07}_{{-0.12}}^{+{0.10}}$ & ${4}_{{-2}}^{+{3}}$ & ${5.35}_{{-0.92}}^{+{1.84}}$ \\
3324040907394753792 & Theia 558 & 1.0 & \textit{Gaia}, PS, SDSS & ${0.79}_{{-0.06}}^{+{0.06}}$ & ${3}_{{-1}}^{+{1}}$ & ${3.99}_{{-0.67}}^{+{1.12}}$ \\
4529222337115434240 & Theia 817 & 1.0 & \textit{Gaia}, PS, SDSS & ${0.69}_{{-0.02}}^{+{0.02}}$ & ${25}_{{-1}}^{+{1}}$ & ${4.47}_{{-1.19}}^{+{1.27}}$ \\
5433483136700686336 & Turner 5 & 0.7 & \textit{Gaia}, \textit{GALEX} & ${0.66}_{{-0.05}}^{+{0.04}}$ & ${212}_{{-12}}^{+{20}}$ & ${4.81}_{{-1.16}}^{+{1.69}}$ \\
4098106821451715584 & UPK 5 & 0.8 & \textit{Gaia}, PS & ${0.50}_{{-0.05}}^{+{0.07}}$ & ${12}_{{-3}}^{+{2}}$ & ${6.10}_{{-1.20}}^{+{1.81}}$ \\
\hline
\end{tabular}
\tablefoot{
WD masses ($M_{\mathrm{WD}}$), 
cooling ages ($t_{\mathrm{cool}}$), and progenitor masses ($M_{\mathrm{i}}$) have been calculated in Sect. 4. $P_{\mathrm{max}}$ represents the maximum value of astrometric membership probability from $P_{\mathrm{3D}}$, $P_{\mathrm{5D}}$, and $P_{\mathrm{HR}}$. \\
\tablefoottext{a}{mentioned in \citet{richer_21} but no $M_{\mathrm{i}}$ calculated},  \tablefoottext{b}{possible double degenerate system \citep{williams_2015}},\tablefoottext{c}{radial velocity possibly inconsistent with OC membership \citep{williams_2015}}, \tablefoottext{d}{approximate parameters derived also in \citet{prisegen_2021} using the GDR2 data}
}
\end{table*}

However, the majority of the OC WD candidates are either novel detections or were previously the subject of only limited study. While $M_{\mathrm{WD}}$ and $t_{\mathrm{cool}}$ can be calculated by interpolating between cooling tracks in the appropriate CMD, a better estimate of these quantities can be obtained by supplementing the \textit{\textit{Gaia}} data by additional high-quality optical and UV photometry. Therefore, we used the \textit{Vizier} service \citep{vizier} to search for additional supplementary SDSS \citep{ahumada_2020}, Pan-STARRS \citep[PS;][]{panstarrs1}, and \textit{\textit{Galaxy Evolution Explorer}} \citep[\textit{\textit{GALEX}};][]{galex1} photometry. Photometric fitting was then done using \texttt{WDPhotTools} \citep{wdphottools}, taking into consideration the group parallaxes and extinctions of the host OCs. We used models from \citet{bedard_2020} assuming pure H atmospheres, and Markov chain Monte Carlo sampling method \textit{emcee} \citep{emcee} was used to find the WD parameters. This yielded bolometric magnitudes and $M_{\mathrm{WD}}$ (together with their uncertainties). We derived $t_{\mathrm{cool}}$ values by using the mappings between the WD parameters using the tool from \citet{cheng_code}. We list the calculated $M_{\mathrm{WD}}$ and $t_{\mathrm{cool}}$ and their uncertainties in Table~\ref{wds_novel}. The listed uncertainties need to be considered as lower limits only since \texttt{WDPhotTools} does not consider uncertainties in the extinction when fitting WD photometry.

After that, $t_{\mathrm{cool}}$ can be subtracted from the total OC age $t_{\mathrm{OC}}$, yielding the lifetime of the progenitor $t_{\mathrm{i}}$. To convert $t_{\mathrm{i}}$ into the initial progenitor mass $M_{\mathrm{i}}$ we adopt the version 1.2S of the PARSEC tracks \citep{bressan_2012} with COLIBRI TP-AGB tracks \citep{marigo_2013}\footnote{Downloaded from \url{http://stev.oapd.inaf.it/cgi-bin/cmd}}. To quantify the uncertainty in the estimate of $M_{\mathrm{i}}$, we conduct 100 Monte Carlo simulations, each time drawing a random value from the distributions of $t_{\mathrm{cool}}$ and $t_{\mathrm{OC}}$. Considering also the OC metallicity, we then obtained mean $M_{\mathrm{i}}$ values and their uncertainties, which are also  listed in Table~\ref{wds_novel}.

Using the derived values of $M_{\mathrm{WD}}$ and $M_{\mathrm{i}}$ for the novel OC WDs and also collating the data from the literature (Table~\ref{wds_literature_parameters}), we construct the IFMR in Fig.~\ref{fig_ifmr}. The IFMR shows several interesting features, which we discuss in this section. However, it also shows a number of probable contaminants well below the IFMR relation; they are probably field WDs that are either spuriously matched to an OC or objects that are the results of atypical or non-isolated evolutionary pathways.

The high-mass end of the IFMR is informative of the maximum mass of an isolated star that leads to the creation of a WD, which is a fundamental astrophysical quantity. This mass delineates the boundary between the stars that undergo a core-collapse SN explosion and those that do not. Therefore, together with the initial mass function (IMF), it controls the SN rate, which then, in turn, controls the formation rate of neutron stars and black holes, chemical enrichment of the ISM and intergalactic medium, dust production, and star formation rate linked to the SN mechanical feedback. The SN rate is quite sensitive to this progenitor mass limit, as the slope of the IMF is relatively steep near its generally adopted value of 8~$\mathrm{M_{\odot}}$. Therefore, a change of even $\sim 1$~$\mathrm{M_{\odot}}$ can substantially alter the expected SN rate. 

Therefore, high-mass WDs that come from isolated stellar evolution in OCs can provide valuable constraints on this progenitor mass limit. However, there is a notable paucity of high-mass WDs in OCs \citep{feelhauer_2003,tremblay_2012,richer_21}. Indeed, the most massive WDs that are OC members only have $M \sim$~1~$\mathrm{M_{\odot}}$, far below the theoretical WD mass limit of $\sim$1.38~$\mathrm{M_{\odot}}$ \citep{nomoto_87}. Moreover, their progenitors do not seem to exceed $\sim$~6~$\mathrm{M_{\odot}}$ \citep[e.g.,][]{cummings_2018,richer_21}. The paucity of high-mass WDs within OCs is noticeable also in this study. 
We detect only six novel WDs with masses $M_{\mathrm{WD}} \gtrsim 1.0$~$\mathrm{M_{\odot}}$ and they do not seem to come from particularly massive progenitors.  Some of this paucity most likely stems from the selection effects. Firstly, with increasing mass WDs get more compact, giving them significantly smaller sizes and therefore lower luminosities, making them more difficult to detect and characterize. Also, high-mass WDs are more likely to be a part of a binary system with a companion that dominates the overall emission of the system. This is due to a general trend of increasing binary fraction with increasing WD stellar progenitor mass \citep[e.g.,][and the references therein]{abt_83,kouwenhoven_2009,moe_2017}. Moreover, high-mass WDs cool relatively rapidly, and the solar neighborhood is deficient in OCs young enough to be currently hosting stars capable of producing massive WDs. Some physical mechanisms can also be at work. As mentioned before, it is possible that WDs are ejected from their parent OCs by velocity kicks imparted during the WD formation, which might be especially relevant for more massive WDs \citep[e.g.,][]{feelhauer_2003,heyl_2007,davis_2008,fregeau_2009,elbadry_2018}. It is sufficient for the kick to be on the order $\sim$1~$\mathrm{km \, s^{-1}}$ for the WD to escape the OC core within a few million years. Especially relevant is the discovery of one such runaway WD originating within the OC Alpha Per. This WD has the progenitor mass of $\sim$8.5~$\mathrm{M_{\odot}}$, placing it at the currently accepted mass limit for the WD formation. Yet, the mass of the WD itself is $\sim$1.2~$\mathrm{M_{\odot}}$, which is still about 0.2~$\mathrm{M_{\odot}}$ below the Chandrasekhar limit \citep{miller_22}. This might hint at the flattening of the IFMR in the region where $M_{\mathrm{i}}>6.0$~$\mathrm{M_{\odot}}$ and moving the new WD formation mass limit by a significant margin to $\gtrsim$12~$\mathrm{M_{\odot}}$. However, more WDs in this mass range must be discovered to put more solid constraints on the IFMR in this region. Yet, moving this limit upward like this would reconcile the observed type II SN rate that seems to be too low if the progenitor mass limit for the WD formation is at $\sim$8~$\mathrm{M_{\odot}}$ \citep{horiuchi_2011}.

Another approach is to address this problem from the opposite side -- by searching for the WD progenitor mass limit by constraining the progenitor mass required for the star to explode in a SN. This can be done in a direct way, by identifying the SN progenitors in the archival images before the SN event and checking for their presence after the SN has faded sufficiently \citep[e.g.,][]{smartt_2015,van_dyk_2017,koplitz_2021,van_dyk_2022}. It is also possible to study the surrounding stellar population in order to get constraints on the SN progenitor mass \citep[e.g.,][]{dias-rodriguez_2018,williams_2018,diaz-rodriguez_2021,kochanek_2022}, as it can be expected that due to the limited lifetime of the high-mass stars, they are still largely clustered together with the other stars of the common origin in the same star-formation event \citep{lada_2003}. This method is viable for both SNe and SN remnants in the Milky Way or other galaxies up to a few megaparsecs where the stellar populations can still be resolved. From these studies, it seems that the minimum progenitor mass required for an SN (and the maximum progenitor mass yielding WDs) is around 7--9~$\mathrm{M_{\odot}}$. 

The paucity of high-mass OC WDs with progenitor masses above 6~$\mathrm{M_{\odot}}$ together with a possible flattening of the IFMR shape toward higher progenitor masses and rather low progenitor masses of some SNe are not straightforward to reconcile. However, this can be resolved if the effects of the stellar binarity are taken into account. Most of the high-mass stars are in a binary and most of them will also experience some kind of close interaction with their companion at some point in their lives \citep[mass stripping, mass accretion, or mergers; e.g.,][]{sana_2012}. These interaction processes play a critical role in the evolution of massive stars because their magnitude, duration, and their timing in the life of the star have a major impact on the mass and structure of the stellar core as the star reaches the terminal phases of its evolution. This then affects the nature and properties of the nascent stellar remnant. Most notably, \citet{podsiadlowski_2004} proposed that the stars with masses $\sim$8--11~$\mathrm{M_{\odot}}$ are expected to explode in an SN if they are the primary component in a compact binary. This can happen if they experience a significant mass loss due to a binary interaction before the onset of the second dredge-up, which normally significantly reduces the core mass of the isolated stars. A second dredge-up phase takes place after the star has reached the asymptotic giant branch phase and after the convective envelope starts reaching into the stellar core, dredging up a significant amount of matter. The core can lose so much mass that it is no longer massive enough for an SN. However, if the stellar envelope is lost before this, the second dredge-up phase cannot take place and the core does not lose mass in this way. Therefore, the isolated stars in this mass range tend to end up as massive WDs, while an identical star within a close binary may explode as an SN and produce a neutron star. For some binaries, the lower limit for the SN explosion may be as low as 6~$\mathrm{M_{\odot}}$ \citep{podsiadlowski_2004}. Therefore, since the IFMR is only constructed considering the isolated stars that do not experience any sort of close binary processes, the relatively high WD progenitor mass limit can be simultaneously reconciled with the lower observed value of the minimum SN progenitor mass limit. In some stars, especially the ones with low metallicities, the rotation speed can also have a significant effect on the mass of the stellar core and envelope, the surface chemical composition, and the stellar wind. These effects are relevant for both isolated and binary stars, but in the latter case, the interplay between the rotational and binary effects can yield additional complexity \citep[e.g.,][]{meynet_2017}. Rapid rotation may cause multiple types of interior mixing processes that do not arise in slowly rotating stars, supplying fresh fuel for nuclear burning into the stellar core. This can prolong the stellar lifetime and also lead to the formation of more massive cores and therefore, more massive WDs \citep[e.g.,][]{cummings_2019}. Thus, in the presence of dispersion of initial rotational velocities, the rotational effects can introduce some additional scatter in the WD progenitor mass limit and in the IFMR in general, which is also apparent in Fig.~\ref{fig_ifmr}.

In the low-mass end, the IFMR exhibits non-monotonic behavior, approximately in the range 1.6~$\mathrm{M_{\odot}} \, \leq M_{\mathrm{i}} \, \leq$~2.1~$\mathrm{M_{\odot}}$ with a peak at $M_{\mathrm{i}} \approx$~1.8~$\mathrm{M_{\odot}}$. \citet{marigo_2020} attribute this kink in the IFMR to the formation of solar metallicity carbon stars on the asymptotic giant branch. The kink is quite unconstrained in its descending phase and near the point where the IFMR curve starts to rise again ($\sim $2--2.5~$\mathrm{M_{\odot}}$). \citet{marigo_2020} list only a single WD in this range from NGC 752. \citet{fragkou_2019} identified PHR 1315-6555, which is a central star of a planetary nebula within OC AL~1 that also falls into this range. Our search yielded two more WDs that are within this range as well -- high-confidence members of Alessi~22 and NGC~752. These WDs provide useful data in this relatively unconstrained mass range of the IFMR. However, their $M_{\mathrm{WD}}$ values are too high and not consistent with the nonlinear IFMR shape in this range as is proposed in \citet{marigo_2020}. Models of \citet{marigo_2021} also predict the existence of a second IFMR kink, located at higher masses in the range of about 4.2~$\mathrm{M_{\odot}} \, \leq M_{\mathrm{i}} \, \leq$~4.8~$\mathrm{M_{\odot}}$, where the IFMR resumes its monotonic increasing trend again starting from $M_{\mathrm{i}} \sim$~5~$\mathrm{M_{\odot}}$. However, the range and shape of the kink are dependent on the exact details of stellar evolution, such as the physics of convection, mixing-length parameter, and mass loss. We find two high-confidence WD members of NGC~2516 and one WD member of NGC~3532 with $M_{\mathrm{i}} \sim$~4.0~$\mathrm{M_{\odot}}$ and with abnormally high masses $M_{\mathrm{WD}} \sim$~1.0~$\mathrm{M_{\odot}}$, lying significantly above the literature IFMR prescriptions. The presence of these WDs in this mass range may indicate a second departure of the monotonic IFMR trend as proposed in \citet{marigo_2021}. However, this deviation can also be attributed to other sources of the IFMR scatter, such as possible past binary interactions \citep[e.g.,][]{temmink_2020}.

\section{Summary and conclusions}
We have studied the WD content of nearby OCs with a primary focus on obtaining tighter constraints on the IFMR, which can be derived semi-empirically by studying the properties of the WDs and their host OCs. Our search for WDs within OCs relied on the astrometric and photometric data provided by the \textit{\textit{Gaia}} mission in its third data release. When such WDs are identified, it is possible to obtain a significantly more precise distance estimate for the WD, which is based on robust statistics that are in turn based on a large number of OC member stars rather than a singular noisy parallax measurement of a single object. A more precise distance obtained in this way translates to more precise knowledge of the fundamental WD properties, most importantly its mass and cooling age. This, in combination with the knowledge of the total age of the OC and its metallicity, can be used to constrain the lifetime of the progenitor by subtracting the cooling age from the total OC age, which can be used to infer the initial mass of the WD progenitor.

After determining the starting sample of the surveyed OCs, and informed by the capabilities of \textit{\textit{Gaia}} and properties of young WDs, we queried the \textit{\textit{Gaia}} archive for the presence of potential WDs associated with these OCs using relatively liberal criteria based on the OC astrometric properties tabulated in \citet{cantat-gaudin_2020}. For each of the OCs that were identified as potentially hosting WDs based on the previous step, we conducted a more detailed search for possible WD members using the HDBSCAN algorithm employed in both five ($\alpha, \delta, \mu_{\alpha^{\star}}, \mu_{\delta}, \varpi$) and three ($\mu_{\alpha^{\star}}, \mu_{\delta}, \varpi$) astrometric dimensions. For each WD candidate detected in this way, we determined its astrometric membership probability using a Monte Carlo approach by drawing random values of its astrometric parameters based on its cataloged astrometric properties, their errors, and covariances. 

For WD candidates with a reasonably high OC astrometric membership probability  ($P_{\mathrm{memb}} \geq 0.5$), we filtered out obvious low-mass or binary outliers unsuitable for the IFMR determination and WD candidates with cooling ages exceeding the total age of their putative parent OCs. We then calculated their masses and cooling ages based on the new estimates of the distances of their putative parent OCs, supplemented with the OC extinction values from the literature. After that, we determined their progenitor lifetimes and progenitor initial masses. Aside from several previously known OC WDs, we characterized 36 WDs with significant OC membership probabilities that had not been characterized in the literature before. These objects would benefit from a spectroscopic follow-up.

The IFMR constructed from the newly characterized and literature OC WDs is consistent with previously published prescriptions, albeit with a large scatter that might be attributed to several extrinsic or intrinsic factors \citep[e.g., a past binary interaction, rotation effects, and metallicity; e.g.,][]{meynet_2017,cummings_2019,temmink_2020}. As in the previous studies, there is still a paucity of high-mass WDs whose progenitor masses even remotely approach the widely adopted upper progenitor mass limit for WD formation of 8~$\mathrm{M_{\odot}}$. This could be caused by the presence of velocity kicks imparted to high-mass WDs upon formation, which often occurs in combination with the presence of a photometrically dominant secondary binary component for many high-mass WDs.\ This makes searching for these systems difficult.

At the present rapid pace of new OC discoveries, one can expect novel OC-WD pairings to be identified in the near future. This would allow us to put even tighter constraints on the IFMR and the research avenues that are connected to it. \textit{\textit{Gaia}}'s recent and upcoming data releases will provide us with an expanded and improved census of nearby coeval stellar populations that are suitable for studying WDs across a wide range of masses, ages, and initial progenitor metallicities. An expanded sample of WDs that are identified to be hosted by these populations, in conjunction with spectroscopic follow-up observations targeting them, is certain to refine our knowledge of the IFMR and the terminal phases of stellar evolution.

\begin{acknowledgements}
  MP is supported by the European Regional Development Fund, project No. ITMS2014+: 313011W085.
  NF acknowledges support from the grant GA{\v C}R 23-07605S.
  This work has made use of data from the European Space Agency (ESA) mission
  \textit{\textit{Gaia}} (\url{https://www.cosmos.esa.int/\textit{Gaia}}), processed by the
  \textit{\textit{Gaia}} Data Processing and Analysis Consortium (DPAC,
  \url{https://www.cosmos.esa.int/web/\textit{Gaia}/dpac/consortium}).  Funding for the
  DPAC has been provided by national institutions, in particular the
  institutions participating in the \textit{\textit{Gaia}} Multilateral Agreement.
  This research has made use of the VizieR catalogue access tool, CDS,
 Strasbourg, France (DOI : 10.26093/cds/vizier). The original description 
 of the VizieR service was published in 2000, A\&AS 143, 23
  This research has made use of the WEBDA database, operated at the Department of Theoretical Physics and Astrophysics of the Masaryk University. We are grateful to the developers and contributors of the many software packages used in this work: Astropy \citep{astropy_2022}, astroquery \citep{astroquery_2019}, astroML \citep{astroML}, HDBSCAN \citep{campello_2013,mcinnes_2017}, Numpy \citep{numpy}, Scipy \citep{2020SciPy-NMeth}, scikit-learn \citep{scikit-learn}, matplotlib \citep{Hunter:2007}, and ezpadova (\url{https://github.com/mfouesneau/ezpadova}).

\end{acknowledgements}

%
%

\bibliographystyle{aa} 
\bibliography{paper_bibliography}

\begin{appendix}

\section{Properties of OCs associated with WD candidates}
\begin{table*}
\caption{Properties of OCs with at least one associated WD candidate as listed in Table~\ref{wds_novel}.}          
\label{table_oc}      
\centering          
\begin{tabular}{l l l l l }     
\hline\hline  
OC & $\varpi_{\mathrm{gr}}$ & $A_{\mathrm{V}}$ & $\log t$ & $[\mathrm{Fe/H}]$ \\
 & ($\mathrm{\mu}as)$ & (mag) & (yr) & (dex) \\
\hline
Alessi 22 & 2821.6$\pm$9.8 & 0.3$\pm$0.13 & 8.8$\pm$0.2 & 0.0$\pm$0.0 \\
Alessi 62 & 1643.6$\pm$9.9 & 0.85$\pm$0.07 & 9.0$\pm$0.0 & 0.1$\pm$0.07 \\
Alessi 94 & 1811.7$\pm$9.7 & 0.12$\pm$0.08 & 8.3$\pm$0.2 & 0.0$\pm$0.0 \\
CWNU 41 & 1371.9$\pm$10.1 & 0.15$\pm$0.1 & 8.8$\pm$0.2 & 0.0$\pm$0.0 \\
HSC 1155 & 3536.1$\pm$9.6 & 0.16$\pm$0.11 & 8.3$\pm$0.3 & 0.0$\pm$0.0 \\
HSC 2304 & 934.3$\pm$10.3 & 0.2$\pm$0.11 & 8.3$\pm$0.2 & 0.0$\pm$0.0 \\
HSC 381 & 3199.8$\pm$8.7 & 0.21$\pm$0.15 & 7.5$\pm$0.3 & 0.0$\pm$0.0 \\
Mamajek 4 & 2251.5$\pm$8.7 & 0.21$\pm$0.03 & 8.8$\pm$0.0 & 0.05$\pm$0.08 \\
NGC 1901 & 2405.6$\pm$9.1 & 0.11$\pm$0.06 & 8.9$\pm$0.0 & -0.04$\pm$0.08 \\
NGC 2358 & 1107.6$\pm$10.2 & 0.37$\pm$0.07 & 8.8$\pm$0.1 & -0.01$\pm$0.09 \\
NGC 2516 & 2455.3$\pm$9.0 & 0.45$\pm$0.05 & 8.4$\pm$0.1 & -0.01$\pm$0.03 \\
NGC 3532 & 2119.3$\pm$9.2 & 0.14$\pm$0.02 & 8.6$\pm$0.0 & 0.05$\pm$0.03 \\
NGC 6633 & 2561.7$\pm$9.5 & 0.77$\pm$0.11 & 8.8$\pm$0.0 & -0.09$\pm$0.06 \\
NGC 752 & 2303.4$\pm$9.2 & 0.16$\pm$0.07 & 9.2$\pm$0.0 & -0.04$\pm$0.05 \\
Ruprecht 147 & 3293.3$\pm$9.4 & 0.29$\pm$0.05 & 9.5$\pm$0.0 & 0.09$\pm$0.05 \\
Stock 2 & 2706.3$\pm$8.7 & 0.79$\pm$0.06 & 8.9$\pm$0.1 & 0.03$\pm$0.05 \\
Theia 248 & 2087.1$\pm$9.1 & 0.61$\pm$0.18 & 8.0$\pm$0.2 & 0.0$\pm$0.0 \\
Theia 558 & 1462.9$\pm$9.7 & 0.65$\pm$0.19 & 8.3$\pm$0.2 & 0.0$\pm$0.0 \\
Theia 817 & 2799.2$\pm$9.2 & 0.25$\pm$0.16 & 8.3$\pm$0.3 & 0.0$\pm$0.0 \\
Turner 5 & 2418.4$\pm$9.2 & 0.23$\pm$0.06 & 8.6$\pm$0.1 & -0.08$\pm$0.04 \\
UPK 5 & 1812.8$\pm$9.7 & 0.74$\pm$0.06 & 8.0$\pm$0.2 & 0.02$\pm$0.05 \\
\hline
\end{tabular}
\tablefoot{OC group parallaxes ($\varpi_{\mathrm{gr}}$) were computed according to the procedure in \citet{maiz_2021}. OC extinctions ($A_{\mathrm{V}}$), total ages ($\log t$), and ($[\mathrm{Fe/H}]$) were taken from \citet{cantat-gaudin_2020, dias_2021, hunt_2023}.}
\end{table*}

\end{appendix}

\clearpage
\onecolumn

\setcounter{table}{0}
\begin{longtable}{llllllll}
\caption{WDs and WD candidates recovered as OC members in the clustering analysis.} 
\label{wds_all} 
\\ \hline\hline
\textit{\textit{Gaia}} DR3 \texttt{source\_id} & associated OC & $P_{\mathrm{3D}}$ & $P_{\mathrm{5D}}$ & $P_{\mathrm{HR}}$ & \texttt{fidelity} & $P_{\mathrm{WD}}$ & $\mathrm{DB_{M}}$ \\
\hline
\endfirsthead
\caption{continued.}\\
\hline\hline
object & associated OC & $P_{\mathrm{3D}}$ & $P_{\mathrm{5D}}$ & $P_{\mathrm{HR}}$ & \texttt{fidelity} & $P_{\mathrm{WD}}$ & $\mathrm{DB_{M}}$ \\
\hline
\endhead
\hline
\endfoot
\hline
\endlastfoot
\object{\textit{Gaia} DR3 2098988107112755712} & \object{ASCC 101} &  & 0.7 &  & 1.0 & 1.0 &  \\
\object{\textit{Gaia} DR3 2879428195013510144} & \object{Alessi 22} &  &  & 1.0 & 1.0 & 1.0 &  \\
\object{\textit{Gaia} DR3 4519349757798439936} & \object{Alessi 62} & 0.8 & 0.5 & 0.6 & 1.0 & 1.0 &  \\
\object{\textit{Gaia} DR3 391939027303287040} & \object{Alessi 94} &  &  & 1.0 & 1.0 & 1.0 &  \\
\object{\textit{Gaia} DR3 386116254240723712} & \object{Alessi 94} &  &  & 1.0 & 1.0 & 1.0 &  \\
\object{\textit{Gaia} DR3 851411295734572416} & \object{CWNU 1095} &  &  & 0.7 & 1.0 & 1.0 &  \\
\object{\textit{Gaia} DR3 5798954758752816512} & \object{CWNU 41} &  &  & 1.0 & 1.0 & 1.0 &  \\
\object{\textit{Gaia} DR3 335525529520188032} & \object{HSC 1155} &  &  & 1.0 & 1.0 & 1.0 &  \\
\object{\textit{Gaia} DR3 3299641271199703680} & \object{HSC 1630} &  &  & 1.0 & 1.0 & 1.0 &  \\
\object{\textit{Gaia} DR3 5257090272981848192} & \object{HSC 2304} &  &  & 0.9 & 1.0 & 1.0 &  \\
\object{\textit{Gaia} DR3 4502736137087916032} & \object{HSC 381} &  &  & 0.9 & 1.0 & 1.0 &  \\
\object{\textit{Gaia} DR3 2060960191793682176} & \object{HSC 601} &  &  & 1.0 & 0.89 & 1.0 &  \\
\object{\textit{Gaia} DR3 4283928577215973120} & \object{IC 4756} & 1.0 & 0.9 & 0.7 & 1.0 & 1.0 &  \\
\object{\textit{Gaia} DR3 6653882460188145152} & \object{Mamajek 4} & 0.5 & 0.3 & 0.7 & 1.0 & 1.0 &  \\
\object{\textit{Gaia} DR3 4008511467191955840} & \object{Melotte 111} & 1.0 & 1.0 & 1.0 & 1.0 & 1.0 & Y \\
\object{\textit{Gaia} DR3 4662157454731788672} & \object{NGC 1901} &  & 0.6 &  & 1.0 & 1.0 &  \\
\object{\textit{Gaia} DR3 4659513404187412736} & \object{NGC 1901} & 0.8 & 0.6 & 0.6 & 1.0 & 1.0 &  \\
\object{\textit{Gaia} DR3 2931807898171448320} & \object{NGC 2358} &  &  & 0.8 & 1.0 &  &  \\
\object{\textit{Gaia} DR3 5538113835730494464} & \object{NGC 2477} &  &  & 0.5 & 1.0 &  &  \\
\object{\textit{Gaia} DR3 5290834387897642624} & \object{NGC 2516} & 1.0 & 0.9 &  & 1.0 & 1.0 & Y \\
\object{\textit{Gaia} DR3 5289447182180342016} & \object{NGC 2516} & 0.9 & 0.6 &  & 1.0 & 1.0 &  \\
\object{\textit{Gaia} DR3 5290719287073728128} & \object{NGC 2516} & 0.9 & 0.7 & 1.0 & 1.0 & 1.0 & Y \\
\object{\textit{Gaia} DR3 5290720695823013376} & \object{NGC 2516} & 0.6 &  &  & 1.0 & 1.0 &  \\
\object{\textit{Gaia} DR3 5294015515555860608} & \object{NGC 2516} & 1.0 & 0.8 &  & 1.0 & 1.0 &  \\
\object{\textit{Gaia} DR3 5290767695648992128} & \object{NGC 2516} & 1.0 & 0.8 & 0.5 & 1.0 & 1.0 &  \\
\object{\textit{Gaia} DR3 664325543977630464} & \object{NGC 2632} &  & 1.0 & 0.9 & 1.0 & 1.0 & Y \\
\object{\textit{Gaia} DR3 661841163095377024} & \object{NGC 2632} &  & 1.0 & 0.6 & 1.0 & 1.0 & Y \\
\object{\textit{Gaia} DR3 662798086105290112} & \object{NGC 2632} &  & 1.0 & 0.5 & 1.0 & 1.0 & Y \\
\object{\textit{Gaia} DR3 662998983199228032} & \object{NGC 2632} &  & 1.0 &  & 1.0 & 1.0 &  \\
\object{\textit{Gaia} DR3 660178942032517760} & \object{NGC 2632} &  & 1.0 & 0.5 & 1.0 & 1.0 & Y \\
\object{\textit{Gaia} DR3 661270898815358720} & \object{NGC 2632} & 0.8 & 1.0 & 0.8 & 1.0 & 1.0 & Y \\
\object{\textit{Gaia} DR3 661010005319096192} & \object{NGC 2632} &  & 1.0 & 0.8 & 1.0 & 1.0 & Y \\
\object{\textit{Gaia} DR3 661297901272035456} & \object{NGC 2632} &  & 1.0 & 0.7 & 1.0 & 1.0 & Y \\
\object{\textit{Gaia} DR3 661311267210542080} & \object{NGC 2632} & 0.5 & 1.0 & 1.0 & 1.0 & 1.0 & Y \\
\object{\textit{Gaia} DR3 665139697978259200} & \object{NGC 2632} &  & 1.0 & 0.6 & 1.0 & 1.0 & Y \\
\object{\textit{Gaia} DR3 661353224747229184} & \object{NGC 2632} &  & 1.0 & 0.9 & 1.0 & 1.0 & Y \\
\object{\textit{Gaia} DR3 659494049367276544} & \object{NGC 2632} &  & 1.0 & 0.6 & 1.0 & 1.0 & Y \\
\object{\textit{Gaia} DR3 5338675689360848256} & \object{NGC 3532} & 0.4 & 0.6 &  & 1.0 &  &  \\
\object{\textit{Gaia} DR3 5337742307052922752} & \object{NGC 3532} & 0.4 & 0.8 & 0.5 & 1.0 & 1.0 &  \\
\object{\textit{Gaia} DR3 5339898655484772864} & \object{NGC 3532} &  & 0.8 &  & 1.0 & 1.0 &  \\
\object{\textit{Gaia} DR3 5340219811654824448} & \object{NGC 3532} & 0.6 & 0.4 &  & 1.0 & 1.0 &  \\
\object{\textit{Gaia} DR3 5340530320614001792} & \object{NGC 3532} &  & 0.5 &  & 1.0 & 1.0 &  \\
\object{\textit{Gaia} DR3 5340165355769599744} & \object{NGC 3532} &  &  & 0.6 & 0.99 & 1.0 &  \\
\object{\textit{Gaia} DR3 5340220262646771712} & \object{NGC 3532} & 0.5 & 1.0 & 0.7 & 1.0 & 1.0 &  \\
\object{\textit{Gaia} DR3 5887666586717940224} & \object{NGC 5822} &  &  & 0.5 & 1.0 & 1.0 &  \\
\object{\textit{Gaia} DR3 4477168746525464064} & \object{NGC 6633} & 0.7 &  &  & 0.78 & 1.0 & Y \\
\object{\textit{Gaia} DR3 4477214475044842368} & \object{NGC 6633} & 0.6 & 0.4 &  & 1.0 & 1.0 & Y \\
\object{\textit{Gaia} DR3 2166915179559503232} & \object{NGC 6991} & 0.7 & 0.4 &  & 1.0 & 1.0 &  \\
\object{\textit{Gaia} DR3 2170776080281869056} & \object{NGC 7092} & 0.9 & 0.8 & 0.8 & 1.0 & 1.0 &  \\
\object{\textit{Gaia} DR3 342523646152426368} & \object{NGC 752} & 0.7 & 0.5 & 0.7 & 1.0 & 1.0 &  \\
\object{\textit{Gaia} DR3 2082008971824158720} & \object{RSG 5} & 0.6 &  & 0.8 & 1.0 & 1.0 &  \\
\object{\textit{Gaia} DR3 4087833117945955840} & \object{Ruprecht 147} & 0.7 & 0.6 &  & 1.0 & 1.0 &  \\
\object{\textit{Gaia} DR3 4087806832745520128} & \object{Ruprecht 147} & 0.8 & 1.0 & 0.6 & 1.0 & 1.0 &  \\
\object{\textit{Gaia} DR3 4088108859141437056} & \object{Ruprecht 147} & 0.9 & 0.8 & 0.7 & 1.0 & 1.0 &  \\
\object{\textit{Gaia} DR3 4183928888026931328} & \object{Ruprecht 147} & 0.9 & 0.8 & 0.6 & 1.0 & 1.0 &  \\
\object{\textit{Gaia} DR3 4183937688413579648} & \object{Ruprecht 147} & 0.9 & 0.9 & 1.0 & 1.0 & 1.0 &  \\
\object{\textit{Gaia} DR3 4183847562828165248} & \object{Ruprecht 147} & 0.8 & 0.8 & 1.0 & 1.0 & 1.0 &  \\
\object{\textit{Gaia} DR3 4183978061110910592} & \object{Ruprecht 147} & 0.6 & 0.6 &  & 0.99 & 1.0 &  \\
\object{\textit{Gaia} DR3 4184148073089506304} & \object{Ruprecht 147} & 0.8 & 0.8 & 0.6 & 1.0 & 1.0 &  \\
\object{\textit{Gaia} DR3 4184169822810795648} & \object{Ruprecht 147} & 0.9 & 0.8 & 0.9 & 1.0 & 1.0 &  \\
\object{\textit{Gaia} DR3 4183926006112672768} & \object{Ruprecht 147} &  &  & 0.9 & 1.0 & 1.0 &  \\
\object{\textit{Gaia} DR3 4183919237232621056} & \object{Ruprecht 147} &  & 0.5 &  & 1.0 & 1.0 &  \\
\object{\textit{Gaia} DR3 4184196073644880000} & \object{Ruprecht 147} & 0.8 & 0.8 & 0.8 & 1.0 & 1.0 &  \\
\object{\textit{Gaia} DR3 1992469104239732096} & \object{Stock 12} & 0.7 & 0.6 & 0.7 & 1.0 & 1.0 & Y \\
\object{\textit{Gaia} DR3 507105143670906624} & \object{Stock 2} &  &  & 0.5 & 0.9 & 1.0 &  \\
\object{\textit{Gaia} DR3 507555904779576064} & \object{Stock 2} & 0.9 & 0.7 &  & 1.0 & 1.0 &  \\
\object{\textit{Gaia} DR3 506862078583709056} & \object{Stock 2} & 0.7 & 0.6 &  & 1.0 & 1.0 &  \\
\object{\textit{Gaia} DR3 506848643933335296} & \object{Stock 2} & 0.6 &  &  & 1.0 & 1.0 &  \\
\object{\textit{Gaia} DR3 3114831641658036608} & \object{Theia 172} &  &  & 0.6 & 0.92 & 1.0 &  \\
\object{\textit{Gaia} DR3 2174431990805230208} & \object{Theia 248} &  &  & 1.0 & 1.0 & 1.0 &  \\
\object{\textit{Gaia} DR3 2170776080281869056} & \object{Theia 517} &  &  & 0.8 & 1.0 & 1.0 &  \\
\object{\textit{Gaia} DR3 3324040907394753792} & \object{Theia 558} &  &  & 1.0 & 1.0 & 1.0 &  \\
\object{\textit{Gaia} DR3 4529222337115434240} & \object{Theia 817} &  &  & 1.0 & 1.0 & 1.0 &  \\
\object{\textit{Gaia} DR3 4530122390454022272} & \object{Theia 817} &  &  & 1.0 & 1.0 & 1.0 &  \\
\object{\textit{Gaia} DR3 5433483136700686336} & \object{Turner 5} &  &  & 0.7 & 0.99 & 1.0 &  \\
\object{\textit{Gaia} DR3 4098106821451715584} & \object{UPK 5} &  &  & 0.8 & 0.78 & 1.0 &  \\
\object{\textit{Gaia} DR3 5914732847840333440} & \object{UPK 624} & 0.7 & 0.6 & 0.9 & 1.0 & 1.0 &  \\
\hline
\end{longtable}
\tablefoot{Columns $P_{\mathrm{5D}}$, $P_{\mathrm{3D}}$, $P_{\mathrm{HR}}$ are the membership probabilities derived in the 5D, 3D, \citet{hunt_2023} membership analysis, respectively. Columns \texttt{fidelity}, $P_{\mathrm{WD}}$, and $\mathrm{DB_{M}}$, denote the astrometric fidelity flag from \citet{rybizky_2022}, probability of the object being a WD from \citet{fusillo_2021}, and if the object is present as a spectroscopically confirmed WD in the Montreal White Dwarf Database \citep{montreal_db}, respectively.}

\end{document}